%
%
%
\def\today{\ifcase\month\or January\or February\or March\or April\or May\or
June\or July\or August\or September\or October\or November\or December\fi
\space\number\day, \number\year}
%
%
\newcount\notenumber

\def\note{\global\advance\notenumber by 1 \footnote{$^{\the\notenumber}$}}
%
%
\newif\ifsectionnumbering
\newcount\eqnumber
\def\cleareqnumber{\eqnumber=0}
\def\numbereq{\global\advance\eqnumber by 1
\ifsectionnumbering\eqno(\the\secnumber.\the\eqnumber)\else\eqno
(\the\eqnumber)\fi}
\def\eqalinno{{\global\advance\eqnumber by 1}
\ifsectionnumbering(\the\secnumber.\the\eqnumber)\else(\the\eqnumber)\fi}
\def\name#1{\ifsectionnumbering\xdef#1{\the\secnumber.\the\eqnumber}
\else\xdef#1{\the\eqnumber}\fi}
\def\nosectionnumbering{\sectionnumberingfalse}
\sectionnumberingtrue
%
%
\newcount\refnumber

\immediate\openout1=refs.tex
\immediate\write1{\noexpand\frenchspacing}
\immediate\write1{\parskip=0pt}
\def\ref#1#2{\global\advance\refnumber by 1%
[\the\refnumber]\xdef#1{\the\refnumber}%
\immediate\write1{\noexpand\item{[#1]}#2}}
\def\tie{\noexpand~}

%
%
\font\twelvebf=cmbx10 scaled \magstep1
\newcount\secnumber

\def\newsection#1.{\ifsectionnumbering\cleareqnumber\else\fi%
	\global\advance\secnumber by 1%
	\bigbreak\bigskip\par%
	\line{\twelvebf \the\secnumber. #1.\hfil}\nobreak\medskip\par\noindent}
%
%
%
\def \sqr#1#2{{\vcenter{\vbox{\hrule height.#2pt
	\hbox{\vrule width.#2pt height#1pt \kern#1pt
		\vrule width.#2pt}
		\hrule height.#2pt}}}}

%
%
%
\newdimen\fullhsize
\def\fiddle{\fullhsize=6.5truein \hsize=3.2truein}
\def\fullline{\hbox to\fullhsize}
\def\mkhdline{\vbox to 0pt{\vskip-22.5pt
	\fullline{\vbox to8.5pt{}\the\headline}\vss}\nointerlineskip}
\def\mkftline{\baselineskip=24pt\fullline{\the\footline}}
\let\lr=L \newbox\leftcolumn
\def\twocolumns{\fiddle
	\output={\if L\lr \global\setbox\leftcolumn=\columnbox
		\global\let\lr=R \else \doubleformat \global\let\lr=L\fi
		\ifnum\outputpenalty>-20000 \else\dosupereject\fi}}
\def\doubleformat{\shipout\vbox{\mkhdline
		\fullline{\box\leftcolumn\hfil\columnbox}
		\mkftline} \advancepageno}
\def\columnbox{\leftline{\pagebody}}
\nosectionnumbering
\magnification=1200
\def\pr#1 {Phys. Rev. {\bf D#1\tie }}
\def\pe#1 {Phys. Rev. {\bf #1\tie}}
\def\pre#1 {Phys. Rep. {\bf #1\tie}}
\def\pl#1 {Phys. Lett. {\bf #1B\tie }}
\def\prl#1 {Phys. Rev. Lett. {\bf #1\tie }}
\def\np#1 {Nucl. Phys. {\bf B#1\tie }}
\def\ap#1 {Ann. Phys. (NY) {\bf #1\tie }}
\def\cmp#1 {Commun. Math. Phys. {\bf #1\tie }}
\def\imp#1 {Int. Jour. Mod. Phys. {\bf A#1\tie }}
\def\mpl#1 {Mod. Phys. Lett. {\bf A#1\tie}}
\def\jhep#1 {JHEP {\bf #1\tie}}
\def\nuo#1 {Nuovo Cimento {\bf B#1\tie}}
\def\rmp#1 {Rev. Mod. Phys. {\bf #1\tie}}
\def\tie{\noexpand~}

\parskip=15pt plus 4pt minus 3pt
\headline{\ifnum \pageno>1\it\hfil Virial expansion in two 
dimensions $\ldots$\else \hfil\fi}
\font\title=cmbx10 scaled\magstep1
\font\tit=cmti10 scaled\magstep1
\footline{\ifnum \pageno>1 \hfil \folio \hfil \else
\hfil\fi}
\raggedbottom


\overfullrule0pt


\rightline{\vbox{\hbox{RU02-6-B}\hbox{hep-th/0307342}}}
\vfill
\centerline{\title The Virial Expansion of a Dilute Bose Gas in Two Dimensions}
\vfill
{\centerline{\title Hai-cang Ren \footnote{$^{\dag}$}
{\rm e-mail: \vtop{\baselineskip12pt
\hbox{ren@theory.rockefeller.edu,}}}}
}
\medskip
\centerline{\tit Physics Department, The Rockefeller
University}
\centerline{\tit 1230 York Avenue, New York, NY
10021-6399}
\vfill
\centerline{\title Abstract}
\bigskip
{\narrower\narrower
In terms of the $s$-wave phase shift of the two-body scattering at thermal wavelength, a systematic perturbative expansion 
of the Virial coefficients is developed for two-dimensional dilute system of bosons in its gaseous phase at low temperature. 
The thermodynamic functions are calculated to the second order of the expansion parameter. The observability of the universal 
low energy limit of the two dimensional phase shift with a quasi-two dimensional atomic gas in an anisotropic trap is 
discussed.
\par}
\vfill\vfill\break


\newsection Introduction.%

The recent advent of laser cooling and atomic trapping techniques makes the physics of a dilute 
quantum gas experimentally accessible, which led to the observation of the Bose-Einstein condensation
of the trapped metallic atoms in three dimensions \ref{\dalfovo}{F. Dalfovo, S. Giorgini, L. P. Pitaevskii and S. Stringari,
\rmp71, 463(1999).}. Theoretical interests on these many-body quantum mechanical system has also 
revived since then. While the Virial expansion was investigated for a three dimensional dilute gas in both 
gaseous phase and condensate phase long ago 
\ref{\lee1}{T. D. Lee and C. N. Yang, Phys. Rev., {\bf 105}, 1119(1957).}, \ref{\lee2}{T. D. Lee and C. N. Yang, 
Phys. Rev. {\bf 113}, 1165(1959); {\bf 116}, 25(1959).} and \ref{\lieb1}{E. Lieb, J. Math. Phys., {\bf 8}, 43(1967).},
a parallel formulation in two dimensions remains to be developed. There are many elegant works 
on 2D bosons concerning the quasi-Bose condensate near the absolute zero 
\ref{\schick}{M. Schick, Phys. Rev. A{\bf 3}, 1067(1971).}
\ref{\popov}{V. N. Popov, Theor. and Math. Phys., 
{\bf 11}, 565(1977).} \ref{\hines}{D. F. Hines, N. E. Frankel and D. J. Mitchell, Phys. Lett. A{\bf 3}, 
1067(1978)} \ref{\fisher}{D. S. Fisher and P. C. Hohenberg, Phys. Rev. {\bf B37}, 4936(1988).} and 
\ref{\Ovchi}{A. A. Ovchinnikov, J. Phys. Condens. Matter {\bf 5}, 8665(1993)}.
The Virial expansion developed in this paper is complementary. As the quasi-two dimensional gas of 
trapped atoms is also experimentally feasible, the result reported in this paper may be brought to a direct 
comparison with the measurements

As is well-known, a perturbative treatment of a dilute Bose gas in two dimensions suffer from two 
difficulties: 1) The scattering amplitude vanishes in the zero energy limit and the Born expansion 
breaks down for a large number of potentials 
\ref{\adhikari}{S. K. Adhikari, Am. J. Phys. {\bf 54}, 362(1986).}\ref{\khuri}{K. Chadan, N. N. Khuri, A. Martin 
and T. T. Wu, Phys. Rev. {\bf D58}, 025014(1998).}. 2) The long range order parameter 
corresponding to the Bose condensate ceases to exist at nonzero temperatures because of the fluctuation of the 
condensate phase \ref{\hohenberg}{P. C. Hohenberg, Phys. Rev., {\bf 158}, 383(1967).}. 
Both difficulties stems from the two dimensional character of the density of states at low energies. 
We shall focus on the first difficulty in this paper.  

The Hamiltonian of a dilute system of interacting bosons is given by
$$H=\sum_{\vec p}(p^2-\mu)b_{\vec p}^\dagger b_{\vec p}+{1\over 2\Omega}
\sum_{\vec p_1,\vec p_2,\vec p_1^\prime,\vec p_2^\prime}\delta_{\vec p_1+\vec p_2,\vec p_1^\prime+\vec p_2^\prime}
<\vec p_1^\prime \vec p_2^\prime|V|\vec p_1 \vec p_2>b_{\vec p_2^\prime}^\dagger
b_{\vec p_1^\prime}^\dagger b_{\vec p_1}b_{\vec p_2}
\numbereq\name{\eqhamil}$$
with $b_{\vec p}$, $b_{\vec p}^\dagger$ the annihilation and creation operators of the bosons in their 
momentum space and $\mu$ the chemical potential. We have chosen the unit of mass such that the mass of a 
boson is ${1\over 2}$. In the rest of this paper, two-body potential $V$ is assumed to be isotropic, 
repulsive and of short range in coordinate space.

The physics of the system depends on the relations among three length scales, the range of the 
interaction, $r_0$, the average inter-particle distance $1/\sqrt{n}$ and the thermal wavelength, 
$$\lambda=\sqrt{4\pi\beta}\numbereq\name{\wavelength}$$ 
with $\beta=(k_BT)^{-1}$. The diluteness is measured by $nr_0^2<<1$ and the classical limit 
corresponds to $n\lambda^2\to 0$. The Virial expansion we shall derive applies to the temperature such that 
$n\lambda^2\sim 1$ when 
quantum coherence becomes significant. The effective expansion parameter is the $s$-wave scattering 
phase shift at the thermal wavelength, which is $O({1\over\ln(\lambda^2 r_0^2)^{-1}})$. Therefore, the 
interaction corrections are far more significant than that of a three dimensional Bose gas at the same 
diluteness.

In the next section, we shall introduce a renormalized two body potential 
as a expansion parameter in the dilute limit, which set up the systematics of the perturbation. 
The corresponding Virial expansion to the second order of the 
renormalized potential will be developed in the section III with typical thermodynamic quantities calculated 
in the section IV to the same order. The convergence of the Virial expansion and its relation to the quasi 
Bose-Einstein condensation will be discussed in the final section.

\newsection The Renormalized Potential.%

The requirement of renormalizing the interaction potential is not unfamiliar in three dimensions with a hard 
sphere potential or more realistic Lennard-Jones potential, for which a straight forward perturbation series 
breaks down because of the singular behavior of the potential at short distance. The natural choice for the 
renormalized potential is the exact two-body scattering amplitude at zero energy (scattering length). What is 
lacking in two dimensions is such 
a natural choice, since the scattering amplitude vanishes at zero energy. This is analogous to perturbation 
theory of the quantum chromodynamics, where the infrared slavery and the asymptotic freedom deprive us of 
a natural scale of the ultraviolet renormalization. A running coupling constant defined at relevant energy scale has 
to be introduced as the expansion parameter. For a two dimensional Bose gas, we need also to introduce a 
running coupling constant, which turns out to be the s-wave phase shift at the thermal wavelength of the bosons.

To replace $<\vec p_1^\prime \vec p_2^\prime|V|\vec p_1 \vec p_2>$ with appropriate renormalized 
quantity at low energies, we focus on the two body sector of the Hamiltonian (\eqhamil) and define
$${\cal V}={1\over 2}(Ve^{-\beta H}e^{\beta H_0}+e^{\beta H_0}e^{-\beta H}V)
\numbereq\name{\eqren}$$
The hermitian operator ${\cal V}$ can be formally expanded according to the power of the potential $V$ and vice 
versa. To the leading order ${\cal V}=V$ and it is straight forward to show that
$$V={\cal V}+{1\over 2}\Big[\int_0^\beta d\tau {\cal V}e^{-\tau H_0}{\cal V}e^{\tau H_0}
+\int_0^\beta d\tau e^{\tau H_0}{\cal V}e^{-\tau H_0}{\cal V}\Big]+O({\cal V}^3)
\numbereq\name{\eqexpan}$$
Sandwiching ${\cal V}$ between two-body states and completing the integral over $\tau$, we find that
$$<\vec p_1^\prime\vec p_2^\prime|V|\vec p_1\vec p_2>=<\vec p_1^\prime\vec p_2^\prime|{\cal V}|\vec p_1\vec p_2>
+{1\over 2}\sum_{\vec k_2,\vec k_2}\Big[{e^{\beta(p_1^2+p_2^2-k_1^2-k_2^2)}-1\over p_1^2+p_2^2-k_1^2-k_2^2}
+{e^{\beta(p_1^{\prime2}+p_2^{\prime2}-k_1^2-k_2^2)}-1\over p_1^{2\prime}+p_2^{2\prime}-k_1^2-k_2^2}\Big]$$
$$\times <\vec p_1^\prime\vec p_2^\prime|{\cal V}|\vec k_1\vec k_2>
<\vec k_1\vec k_2|{\cal V}|\vec p_1\vec p_2>+O({\cal V}^3).
\numbereq\name{\eqvb}$$
As we shall see, the matrix element $<\vec p_1^\prime\vec p_2^\prime|{\cal V}|\vec p_1\vec p_2>$
vanishes in the limit $\beta\to\infty$ and is a proper choice of the renormalized potential at low 
temperature. Furthermore, replacing $V$ by its formal expansion in ${\cal V}$, the systematics to all
orders is restored. In terms of the two-body matrix element of ${\cal V}$, the Hamiltonian of the system, (\eqhamil)
becomes
$$H=\sum_{\vec p}(p^2-\mu)b_{\vec p}^\dagger b_{\vec p}+{1\over 2\Omega}
\sum_{\vec p_1,\vec p_2,\vec p_1^\prime,\vec p_2^\prime}\delta_{\vec p_1+\vec p_2,\vec p_1^\prime+\vec p_2^\prime}
<\vec p_1^\prime \vec p_2^\prime|{\cal V}|\vec p_1 \vec p_2>b_{\vec p_2^\prime}^\dagger
b_{\vec p_1^\prime}^\dagger b_{\vec p_1}b_{\vec p_2}$$
$$+{1\over 2\Omega}\sum_{\vec p_1,\vec p_2,\vec p_1^\prime,\vec p_2^\prime}
\delta_{\vec p_1+\vec p_2,\vec p_1^\prime+\vec p_2^\prime}<\vec p_1^\prime \vec p_2^\prime|
{\cal V}-V|\vec p_1 \vec p_2>b_{\vec p_2^\prime}^\dagger b_{\vec p_1^\prime}^\dagger b_{\vec p_1}b_{\vec p_2},
\numbereq\name{\eqhamilr}$$
where the last term is understood as an power series of ${\cal V}$ starting from the order ${\cal V}^2$, like 
the renormalization counter terms in  relativistic field theories. A similar method of renormalization was developed in 
the context of a lattice gas with on-site exclusion. \ref{\ren}{R. Friedberg, T. D. Lee and Hai-cang Ren, Phys. Rev. {\bf B50}, 10190(1994.}

The matrix element (\eqren) is related to the binary kernel of Lee and Yang [\lee2] through
$$<\vec p_1^\prime\vec p_2^\prime|{\cal V}|\vec p_1\vec p_2>=
-{1\over 2}\Big[<\vec p_1^\prime\vec p_2^\prime|B(\beta)|\vec p_1\vec p_2>e^{\beta(p_1^2+p_2^2)}
+<\vec p_1\vec p_2|B(\beta)|\vec p_1^\prime\vec p_2^\prime>^*e^{\beta(p_1^{\prime2}+p_2^{\prime2})},
\numbereq\name{\eqLeeYang}$$
and the Virial expansion developed here for a 2D Bose gas is parallel to that of Lee and Yang for a 3D
hard sphere gas. 

In terms of the total momentum $\vec P=\vec p_1+\vec p_2$, $\vec P^\prime=\vec p_1^\prime+\vec p_2^\prime$ and the 
relative momentum $\vec p={1\over 2}(\vec p_1-\vec p_2)$, $\vec p^\prime={1\over 2}(\vec p_1^\prime-\vec p_2^\prime)$, 
we have $$<\vec p_1^\prime\vec p_2^\prime|{\cal V}|\vec p_1\vec p_2>
={1\over 2}\delta_{\vec P^\prime,\vec P}[e^{2\beta p^2}<\vec p^\prime|Ve^{-\beta h}|\vec p>
+e^{2\beta p^{\prime2}}<\vec p^\prime|e^{-\beta h}V|\vec p>],
\numbereq\name{\eqbinary}$$
where $$h=-2\nabla_r^2+V(\vec r)\numbereq\name{\potential}$$ is the Hamiltonian of a potential scattering problem with 
$\vec r$ the relative coordinate. Inserting the complete set of eigenstates of $h$, i.e., $h|n>=E_n|n>$, we have
$$<\vec p^\prime|Ve^{-\beta h}|\vec p>=\sum_ne^{-\beta E_n}<\vec p^\prime|V|n><n|\vec p>
\numbereq\name{\formal}$$
and $<\vec p^\prime|e^{-\beta h}V|\vec p>=<\vec p|Ve^{-\beta h}|\vec p^\prime>^*$. For an isotropic 
potential, $|n>$ is specified by the azimuthal quantum number, $m$, and the radial momentum, $k$, with $E=2k^2$. The 
corresponding wave function 
$$<\vec r|k,m>=N_{km}{1\over\sqrt{2\pi}}e^{im\phi}u_m(k|r)
\numbereq\name{\wave}$$ where the radial wave function $u_m(k|r)$ approaches asymptotically
$$u_m(r|k)\simeq \sqrt{{2\over\pi kr}}\cos\Big[kr-{m\pi\over 2}-{\pi\over 4}+\delta_m(k)\Big]
\numbereq\name{\asympt}$$ for $kr>>1$ with $\delta_m(k)$ the phase shift and $N_{km}$ is a normalization constant. 
The partial-wave expansion of (\formal) reads
$$<\vec p^\prime|Ve^{-\beta h}|\vec p>={1\over\Omega}V_0(p^\prime, p)+{2\over\Omega}\sum_{m=1}^\infty 
 V_m(p^\prime, p)\cos m\Phi,
\numbereq\name{\partialw}$$
where 
$$V_m(p^\prime,p)=2\pi\sum_kN_{km}^2e^{-2\beta k^2}\int_0^\infty dr^\prime r^\prime J_m(p^\prime r^\prime)V(r^\prime)
u_m(k|r^\prime)\int_0^\infty drru_m(k|r)J_m(pr)
\numbereq\name{\bessel}$$
with $\Phi$ the angle between $\vec p$ and $\vec p^\prime$, and $J_m(z)$ the Bessel function. The infinite volume limit 
of the integral over $r$ and the sum over $k$ in (\bessel) have to be evaluated carefully. We defer the details to the 
appendix A and quote only the result here,
$$V_m(p^\prime,p)=-4e^{-2\beta p^2}\Big({p^\prime\over p}\Big)^m\sin 2\delta_m(p)
+\pi{\cal P}\int_0^\infty dkke^{-2\beta k^2}{v_m(p^\prime,k)v_m(p,k)\over k^2-p^2}
\numbereq\name{\eqrel}$$
with $$v_m(p,k)=\int_0^\infty drrJ_m(pr)V(r)u_m(k|r),\numbereq\name{\eqvm}$$ 
$$v_m(p,p)=-{4\over\pi}\sin\delta_m(p).\numbereq\name{\eqvm2}$$ and ${\cal P}$ the principal value of the 
integral. For a short-range potential, the wave function $u_m(k|r)$ normalized according to (\asympt) takes the form
$$u_m(k|r)\simeq {\sin\delta_m(k)\over k^{2m}}f_m(r)\numbereq\name{\lowk}$$
in the limit $kr\to 0$, with $f_m(r)$ independent of $k$. The phase shifts display the following low energy behavior
$$\delta_0(k)\simeq {\pi\over 2\ln ka}
\numbereq\name{\phase0}$$
and
$$\delta_m(k)\sim k^{2m},
\numbereq\name{\phasem}$$
where $a$ is the $s$-wave scattering length in two dimensions. For a hard sphere potential, 
$a={1\over 2}e^\gamma r_0$ with $r_0$ the radius of the sphere and $\gamma$ the Euler constant.
But the scattering length and the range of the potential may not be comparable in general. The relevant length for the 
Virial expansion developed in this paper is the scattering length $a$. 
It follows from (\eqvm2) and (\lowk) that 
$$v_m(p,k)=-{4\over\pi}\Big({p\over k}\Big)^m\sin\delta_m(k)\numbereq\name{\eqvm3}$$ for $k$ and $p$ both small, and 
the low energy and low temperature approximation of (\eqrel) reads
$$V_m(p^\prime,p)\simeq-4e^{-2\beta p^2}\Big({p^\prime\over p}\Big)^m\sin 2\delta_m(p)
+{16\over\pi}p^{\prime m}p^m{\cal P}\int_0^\infty dkk^{-2m+1}e^{-2\beta k^2}{\sin^2\delta_m(k)\over k^2-p^2}
\numbereq\name{\eqapprox}$$
  
The eq. (\phase0) was proved rigorously by Chan, Khuri, Martin and Wu [\khuri] for a general class of potentials that fall 
off faster than ${1\over r^2\ln r}$ for $r\to\infty$ . They also proved that for the same class of potentials, 
the correction to the corresponding function $f_0(r)$ of is of the order of $k^2$. As this involves only the long wavelength 
limit of the scattering, their conclusion can also generalized to the repulsive potential that becomes singular as $r\to 0$. 
Therefore, all the logarithmic dependence on the range of the 
interaction of, $\ln a$, is absorbed in the $s$-wave phase shift through (\eqapprox). 
This way the $s$-wave channel dominant over all other partial wave channels to all orders of a low energy expansion, different from 
three dimensions.  The universality found in [\khuri] is highlighted in the low temperature thermodynamics of the 
2D system. In what follows, we shall suppress the subscript ``0'' for the $s$-wave and introduce 
a running coupling constant at the scale of the thermal wavelength, 
$$\alpha\equiv{1\over\ln{\lambda^2\over 2\pi a^2}-\gamma},
\numbereq\name{\run}$$
we have $\alpha\simeq{1\over\pi}\delta({\sqrt{2\pi}e^{\gamma\over 2}\over\lambda})$ and 
the $s$-wave phase shift becomes
$$\delta=-\pi\alpha+\pi\alpha^2\Big(\ln{k^2\lambda^2\over 2\pi}+\gamma\Big)+...
\numbereq\name{\eqsw}$$
AS we shall see, the choice of the constant pertaining to the logarithm of (\run) is to make the second Virial coefficient 
free from $O(\alpha^2)$ corrections.
It follows from (\formal), (\eqrel) and (\eqsw) that the renormalized potential at low energies reads
$$<\vec p_1^\prime\vec p_2^\prime|{\cal V}|\vec p_1\vec p_2>={\delta_{\vec P,\vec P^\prime}\over\Omega}\Big[
8\pi\alpha-8\pi\alpha^2\Big(\ln{pp^\prime\lambda^2\over 2\pi}+\gamma\Big)
+16\pi\alpha^2{\cal P}\int_0^\infty dkk{e^{2\beta (p^2-k^2)}\over k^2-p^2}+...\Big],\numbereq\name{\ren}$$
and this expression will be applied extensively in the subsequent sections. 

The dependence of the dimensionless running coupling constant (\run) on $T$ and $a$ is completely analogous to that of 
a relativistic field theory of zero masses in 3D with $T$ corresponding to the renormalization energy scale and $a$ 
the ultraviolet cutoff.

\newsection Virial Expansion.%

The thermodynamics of a uniform gas is determined completely by its equation of state, usually expressed in the form 
of the Virial expansion,
$${p\over k_BT}=\sum_{l=1}^\infty b_lz^l
\numbereq\name{\pressure}$$
and
$$n={\partial\over\partial\ln z}\Big({p\over k_BT}\Big)_T
\numbereq\name{\density}$$
with $n$ the number density and $z=e^{\beta\mu}$ the fugacity. The $l$-th Virial coefficient, $b_l$ is determined by 
the quantum mechanics of $l$ particles, the exact solution of which is in general unavailable for $l>2$. For 
an ideal Bose gas in two dimensions, $b_l={1\over\lambda^2l^2}$. On writing 
$$b_l={1\over\lambda^2l^2}+b_l^\prime,
\numbereq\name{\viril}$$
we have 
$${p\over k_BT}={1\over\lambda^2}g_2(z)+\Gamma
\numbereq\name{\pert}$$
with $g_2(z)=\sum_{l=1}^\infty{z^l\over l^2}$ and $\Gamma\equiv\sum_{l=1}^\infty b_l^\prime z^l$. The 
perturbative expansion of $\Gamma$ is represented by thermal diagrams. 

The thermal diagrams of $\Gamma$ to the third order in $\alpha$ is shown in Fig. 1, where a solid line represents 
a boson propagator, ${i\over i\omega_n-p^2+\mu}$ with $\omega_n$ the Matsubara energy of the boson, 
a solid circle vertex is associated to the factor $<\vec p_1^\prime\vec p_2^\prime|{\cal V}|\vec p_1\vec p_2>$ with 
$\vec p_1$, $\vec p_2$ ($\vec p_1^\prime$, $\vec p_2^\prime$) the incoming (outgoing) momenta and an open 
circle vertex denotes
$<\vec p_1^\prime\vec p_2^\prime|V-{\cal V}|\vec p_1\vec p_2>$, the analog of the renormalization counter term in 
a relativistic field theory. In what follows, we shall calculate the $\Gamma$ to the order $\alpha^2$, which is 
one order beyond the mean field approximation.

On writing 
$$\Gamma=\Gamma_a+\Gamma_b+\Gamma_c+\Gamma_d+...\numbereq\name{\diagram}$$
with $\Gamma_a$, $\Gamma_b$, $\Gamma_c$ and $\Gamma_d$ standing for the contribution from the first four of 
the diagrams in Fig. 1 in a sequential order, we have
$$\Gamma_a=-{1\over\beta^2}\sum_{\omega_1,\omega_2}{1\over(i\omega_1-p_1^2+\mu)(i\omega_2-p_2+\mu)}
\int{d^2\vec p_1\over (2\pi)^2}\int{d^2\vec p_2\over (2\pi)^2}<\vec p_1\vec p_2|{\cal V}|\vec p_1\vec p_2>
n(\vec p_1)n(\vec p_2)$$
$$=-\beta\int{d^2\vec p_1\over (2\pi)^2}\int{d^2\vec p_2\over (2\pi)^2}<\vec p_1\vec p_2|{\cal V}|\vec p_1\vec p_2>
n(\vec p_1)n(\vec p_2)\numbereq\name{\first}$$
with $$n(\vec p)={ze^{-\beta p^2}\over 1-ze^{-\beta p^2}}
=\sum_{l=1}^\infty z^le^{-l\beta p^2}\numbereq\name{\taylor}$$ and $z=e^{\beta\mu}$ the fugacity. In accordance with 
the expansion (\ren), we have 
$$\Gamma_a=\Gamma_a^\prime+\Gamma_a^{\prime\prime}\numbereq\name{\decomp}$$
with 
$$\Gamma_a^\prime=-8\pi\alpha\beta\int{d^2\vec p_1\over (2\pi)^2}\int{d^2\vec p_2\over (2\pi)^2}n(\vec p_1)n(\vec p_2)
\Big[1-\alpha\Big(\ln{p^2\lambda^2\over 2\pi^2}+\gamma\Big)\Big]
\numbereq\name{\first1}$$ 
and 
$$\Gamma_a^{\prime\prime}=16\pi\alpha^2\beta\int{d^2\vec p_1\over (2\pi)^2}\int{d^2\vec p_2\over (2\pi)^2}
n(\vec p_1)n(\vec p_2){\cal P}\int_0^\infty dkk{e^{\beta(p^2-k^2)}\over p^2-k^2}.
\numbereq\name{\first2}$$
Using Taylor expansion of (\taylor) and integrating over $\vec p_1$ and $\vec p_2$, we end up with
$$\Gamma_a^\prime= -{2\alpha\over\lambda^2}\Big[\ln^2{1\over 1-z}+\alpha D(z)+O(\alpha^2)\Big]
\numbereq\name{\K_U}$$
with
$$D(z)=\sum_{r,s=1}^\infty{z^{r+s}\over rs}\ln{2rs\over r+s}.$$
Similarly, we have
$$\Gamma_b={128\pi^2\alpha^2\over \beta^2}\int{d^2\vec p_1\over (2\pi)^2}\int{d^2\vec p_2\over (2\pi)^2}
\int{d^2\vec p_3\over (2\pi)^2}\sum_{\omega_1,\omega_2,\omega_3}{1\over(i\omega_1-p_1^2+\mu)^2
(i\omega_2-p_1^2+\mu)(i\omega_3-p_3^2+\mu)}$$
$$=128\pi^2\alpha^2\beta^2\int{d^2\vec p_1\over (2\pi)^2}\int{d^2\vec p_2\over (2\pi)^2}\int{d^2\vec p_3\over (2\pi)^2}
n(\vec p_1)[n(\vec p_1)+1]n(\vec p_2)n(\vec p_3)={8\alpha^2\over\lambda^2}{z\over(1-z)^2}\ln^2{1\over 1-z}
\numbereq\name{\second}$$
Finally
$$\Gamma_c=32\pi^2\alpha^2\beta\int{d^2\vec p_1\over (2\pi)^2}\int{d^2\vec p_2\over (2\pi)^2}
\int{d^2\vec p_3\over (2\pi)^2}\int{d^2\vec p_4\over (2\pi)^2}(2\pi)^2\delta^2(\vec p_1+\vec p_2-\vec p_3-\vec p_4)$$ 
$$\times\sum_{\omega_1,\omega_2\omega_3}{1\over
(i\omega_1-p_1^2+\mu)(i\omega_2-p_2^2+\mu)(i\omega_3-p_3^2+\mu)[i(\omega_1+\omega_2-\omega_3)-p_1^2+\mu]}
$$ $$={32\pi^2\alpha^2\beta\over z^2}\int{d^2\vec p_1\over (2\pi)^2}\int{d^2\vec p_2\over (2\pi)^2}\int{d^2\vec p_3\over (2\pi)^2}
\int{d^2\vec p_4\over (2\pi)^2}(2\pi)^2\delta^2(\vec p_1+\vec p_2-\vec p_3-\vec p_4)$$
$$\times {e^{\beta(p_1^2+p_2^2)}-e^{\beta(p_3^2+p_4^2)}\over p_1^2+p_2^2-p_3^2-p_4^2}
n(\vec p_1)n(\vec p_2)n(\vec p_3)n(\vec p_4),
\numbereq\name{\third}$$
which represents a genuine three-body scattering, and
$$\Gamma_d=-64\pi^2\alpha^2\beta\int{d^2\vec p_1\over (2\pi)^2}\int{d^2\vec p_2\over (2\pi)^2}
\int{d^2\vec p_3\over (2\pi)^2}\int{d^2\vec p_4\over (2\pi)^2}(2\pi)^2\delta^2(\vec p_1+\vec p_2-\vec p_3-\vec p_4)$$
$$\times {e^{\beta(p_1^2+p_2^2-p_3^2-p_4^2)}-1\over p_1^2+p_2^2-p_3^2-p_4^2}n(\vec p_1)n(\vec p_2),
\numbereq\name{\forth}$$
Things are greatly simplified to the order $O(\alpha^2)$ by forming the following combination
$$\Gamma_c^\prime\equiv \Gamma_a^{\prime\prime}+\Gamma_c+\Gamma_d$$ $$=-128\pi^2\alpha^2\beta
{\cal P}\int{d^2\vec p_1\over (2\pi)^2}\int{d^2\vec p_2\over (2\pi)^2}\int{d^2\vec p_3\over (2\pi)^2}
\int{d^2\vec p_4\over (2\pi)^2}(2\pi)^2\delta^2(\vec p_1+\vec p_2-\vec p_3-\vec p_4)
{n(\vec p_1)n(\vec p_2)n(\vec p_3)\over p_1^2+p_2^2-p_3^2-p_4^2}.
\numbereq\name{\final}$$
This expression is then simplified in four steps: 

1) Expanding $n(\vec p_i)$ according to the power of $z$;

2) Using the property of the principal value,
$${\cal P}{1\over p_1^2+p_2^2-p_3^2-p_4^2}={\rm{Re}}{1\over p_1^2+p_2^2-p_3^2-p_4^2+i0^+}
={\rm{Im}}\int_0^\infty dx e^{i(p_1^2+p_2^2-p_3^2-p_4^2+i0^+)x}
\numbereq\name{\principal}$$
and the Fourier transformation of the delta function,
$$\delta^2(\vec p_1+\vec p_2-\vec p_3-\vec p_4)=\int{d^2\vec\rho\over(2\pi)^2}
e^{i(\vec p_1+\vec p_2-\vec p_3-\vec p_4)\cdot\vec\rho};
\numbereq\name{\deltaf}$$ 

3) Carrying out the Gauss integration over $\vec p's$ and then the Gauss 
integration over $\vec\rho$ for each term of the power series in $z$; 

4) Carrying out the elementary integral over $x$. The final result 
reads
$$\Gamma_c^\prime={4\alpha^2\over\lambda^2}F(z)\numbereq\name{\res}$$
with 
$$F(z)=\sum_{r,s,t=1}^\infty{z^{r+s+t}\over\sqrt{rs(r+t)(s+t)}}\ln{\sqrt{(r+t)(s+t)}+\sqrt{rs}
\over\sqrt{(r+t)(s+t)}-\sqrt{rs}}.$$

Collecting above results, we obtain the Virial expansion of a dilute Bose gas to the second 
order of the interaction, 
$${p\over k_BT}={1\over\lambda^2}g_2(z)-{2\alpha\over \lambda^2}\ln^2{1\over 1-z}
+{2\alpha^2\over\lambda^2}\Big[{4z\over 1-z}\ln^2{1\over 1-z}+2\phi(z)\Big]+O(\alpha^3),
\numbereq\name{\eqstate}$$ with 
$$\phi(z)\equiv F(z)+{1\over 2}D(z),\numbereq\name{\phifunc}$$
and the corresponding number density is given by
$$n={1\over\lambda^2}\ln{1\over 1-z}-{4\alpha\over \lambda^2}{z\over 1-z}\ln{1\over 1-z}
+{2\alpha^2\over\lambda^2}{d\over d\ln z}\Big[{4z\over 1-z}\ln^2{1\over 1-z}+2\phi(z)\Big]+O(\alpha^3).
\numbereq\name{\eqdense}$$
We notice that the order $\alpha^2$ term start with the third power of $z$, which may be viewed as a 
criteria to fix the constant part pertaining the logarithm of the running coupling constant
(\run). Furthermore, the coefficient of $z^2$ agrees with the result obtained with the classical formula of
Beth and Uhlenbeck \ref{\beth}{E. Beth and G. E. Uhlenbeck, Physica, {\bf 4}, 915 (1937).}. 
The asymptotic behavior of $D(z)$ and $F(z)$ as $z\to 1^-$ is analyzed in the appendix B.

\newsection Thermodynamical Functions.%

Among experimental observables of a two dimensional Bose gas are the homogeneous thermodynamical functions,
which follow readily from the Virial expansion of the last section. For the sake of 
clarity of notations, all thermodynamic functions in their ideal gas limit will carry the superscript '(0)'.

Inverting (\eqdense), an expression of 
the fugacity $z$ in terms of the density $n$ is obtained to the order $\alpha^2$, 
$$\ln z=\ln z^{(0)}+4\alpha\xi-{4\alpha^2\over e^{\xi}-1}{d\phi\over d\ln z}\Big|_{z=z^{(0)}}\numbereq\name{\eqfug}$$
with $\xi\equiv n\lambda^2$ and $z^{(0)}=1-e^{-\xi}$, the fugacity of an ideal Bose gas in two dimensions. On
substituting (\eqfug) back to (\eqstate), we obtain the equation of state to the order $\alpha^2$, 
$$p=p^{(0)}+8\pi\alpha n^2+{16\pi\over\lambda^4}\alpha^2\Big[\phi(1-e^{-\xi})-{\xi\over e^{\xi}-1}{d\phi\over d\ln z}
\Big|_{z=1-e^{-\xi}}\Big],\numbereq\name{\eqpres}$$ where 
$$p^{(0)}={4\pi\over\lambda^4}g_2(1-e^{-\xi})={4\pi\over\lambda^4}\Big[\xi-{\xi^2\over 4}
+\sum_{m=1}^\infty{(-)^{m-1}B_m\over(2m+1)!}\xi^{2m+1}\Big]\numbereq\name{\ideal}$$ 
is the equation of state for an ideal gas with $B_m$ the $m$-th Bernoullian number.

The Helmholtz free energy per unit volume is obtained through the formula
$$f=\mu n-p=f^{(0)}+8\pi\alpha n^2-{16\pi\alpha^2\over\lambda^4}\phi(1-e^{-\xi})
\numbereq\name{\eqfree}$$ 
The entropy per unit volume is given by 
$$s=\Big({\partial p\over\partial T}\Big)_\mu=s^{(0)}-{16\pi\alpha^2\over \lambda^4T}\Big[
\xi^2-2\phi(1-e^{-\xi})+{\xi\over e^{\xi}-1}{d\phi\over d\ln z}\Big|_{z=1-e^{-\xi}}\Big]\numbereq\name{\eqentropy}$$ 
and the correction to the ideal limit is of the $\alpha^2$. It follows from (\eqfree) 
and (\eqentropy) that the internal energy per unit volume is
$$u=f+Ts=u^{(0)}+8\pi\alpha n^2-{16\pi\alpha^2\over\lambda^4}\Big[\xi^2-\phi(1-e^{-\xi})
+{\xi\over e^{\xi}-1}{d\phi\over d\ln z}\Big|_{z=1-e^{-\xi}}\Big]
\numbereq\name{\eqenergy}$$

The specific heat at a constant volume(area) reads
$$c_V=T\Big({\partial s\over\partial T}\Big)_n=c_V^{(0)}+{4\alpha^2k_B\over\lambda^2}\Big[\xi^2+2\phi(z)
-{\xi(2e^{\xi}-2+\xi e^{\xi})\over (e^{\xi}-1)^2}{d\phi\over d\ln z}
+{\xi^2\over(e^{\xi}-1)^2}{d^2\phi\over d(\ln z)^2}\Big]\Big|_{z=1-e^{-\xi}},\numbereq\name{\spec}$$
and the leading order contribution of the interaction is proportional to $\alpha^2$.
The isothermal compressibility, $\kappa_T$ can be calculated readily from (\eqpres)  with the results 
$$\kappa_T=-\Omega\Big({\partial p\over\partial\Omega}\Big)_T=\kappa_T^{(0)}+16\pi\alpha n^2
+{16\pi\alpha^2 n^2\over (e^\xi-1)^2}\Big[e^\xi{d\phi\over d\ln z}-{d^2\phi\over d(\ln z)^2}\Big]
\Big|_{z=1-e^{-\xi}}.\numbereq\name{\eqcompT}$$ 
Employing the thermodynamic relationship 
$$\kappa_S-\kappa_T=-{n^2T\over c_V}\Big({\partial p\over\partial T}\Big)_n\Big[{\partial\over\partial n}
\Big({s\over n}\Big)\Big]_T\numbereq\name{\relation}$$
we find the adiabatic compressibility 
$$\kappa_S=-\Omega\Big({\partial p\over\partial\Omega}\Big)_S=\kappa_S^{(0)}+16\pi\alpha n^2
+16\pi\alpha^2n^2+{32\pi\alpha^2\over\lambda^4}\Big[\phi(z)-{\xi\over e^\xi-1}{d\phi\over d\ln z}\Big]
\Big|_{z=1-e^{-\xi}},\numbereq\name{\eqcompS}$$
which can be directly measured through the sound speed in the Bose gas,
$$v=\sqrt{{2\kappa_S\over n}}.\numbereq\name{\sound}$$

\newsection Concluding Remarks.%

In previous sections, we have developed a diagrammatic approach to the Virial expansion of a dilute 
Bose gas with a repulsive interaction of a short range, valid in the region where $n\lambda^2\sim 1$ 
and $\lambda>>a$. Before concluding the paper, we shall remark on its implication on the quasi Bose-Einstein 
condensation and its applicability to the realistic system of an quasi-2D gas of alkaline atoms in a strongly 
anisotropic trap. 

All coefficients of the Virial expansion (\eqstate) are singular at $z=1$, and the singularity gets 
stronger at higher orders. A simple power counting argument shows that the coefficient of $\alpha^N$ diverges 
as $(1-z)^{1-N}$ as $z\to 1^-$, up to some powers of $\ln{1\over 1-z}$, as is bore out in the  explicit form of 
(\eqstate) to $\alpha^2$ (See appendix B for 
the singularities of $D(z)$ and $F(z)$). Therefore 
the reliability of the expansion requires that 
$$\alpha << 1-z\numbereq\name{\valid}$$
Using the formula (\run) for $\alpha$ and the ideal gas limit of $z$, the condition (\valid) becomes  
$$T>>{4\pi n\over\ln\ln{1\over 2\pi na^2}},\numbereq\name{\critical}$$ which is consistent with the formula 
of the transition temperature to the superfluid phase, obtained in [\popov] and [\fisher], under a mean 
field approximation. It will be interesting to see whether the resummation of diagrams developed in 
\ref{\lee3}{T. D. Lee and C. N. Yang, Phys. Rev., {\bf 117}, 12(1960).} for a 3D hard sphere Bose gas can be 
applied here to extract the ground state energy beyond the mean-field approximation and to compare with the results 
in [\hines] and [\Ovchi]. At this stage, we
only remark on that the order $\alpha$ correction to the internal energy is twice of the rigorous result of 
Lieb and Yngvason \ref{\lieb}{E. H. Lieb and J. Yngvason, J. Stat. Phys., {\bf 103}, 509(2001).} for the 
ground state energy for the thermal wavelength comparable with the inter-boson 
distance. The factor two comes from the exchange energy which is absent at zero temperature because of the Bose 
condensate. For a point-like repulsion corresponding to the leading order term of (\ren), the exchange energy is 
equal to the classical energy and thereby doubles the interaction energy under the mean-field approximation. 
A similar effect was observed in three dimensions [\lee3]\footnote{$^{\dag}$}
{\rm Another way to understand the factor two is to consider the expectation value of the potential energy of a 
point-like repulsion with respect to a state of ${\cal N}(=n\Omega)$ free bosons, i. e. 
$E\equiv <|{V\over 2\Omega}\sum_{\vec p_1,\vec p_2,\vec p_1^\prime,\vec p_2^\prime}
\delta_{\vec p_1+\vec p_2,\vec p_1^\prime+\vec p_2^\prime}b_{\vec p_2^\prime}^\dagger
b_{\vec p_1^\prime}^\dagger b_{\vec p_1}b_{\vec p_2}|>$ with $|>$ a product of single particle states. 
The summation can be broken into three nonvanishing 
terms: $E={V\over 2\Omega}[\sum_{\vec p_1,\vec p_2}<|b_{\vec p_2}^\dagger
b_{\vec p_1}^\dagger b_{\vec p_1}b_{\vec p_2}|>+\sum_{\vec p_1,\vec p_2}<|b_{\vec p_1}^\dagger
b_{\vec p_2}^\dagger b_{\vec p_1}b_{\vec p_2}|>-\sum_{\vec p}<|b_{\vec p}^\dagger
b_{\vec p}^\dagger b_{\vec p}b_{\vec p}|>]$. In the limit of an infinite volume, $\Omega\to\infty$, each of 
the first and second terms gives rise to ${\cal N}^2$. If all bosons occupy one momentum level (Bose condensate), 
the third term contributes
$-{\cal N}^2$, making the sum ${E_0\over\Omega}={1\over 2}Vn^2$. If none of the levels are macroscopically 
occupied, the third term is $O({\cal N})$ and will not contribute to the infinite volume limit. 
We have then ${E_0\over\Omega}=Vn^2$.}

The Virial expansion in two dimensions is characterized by the logarithm in the denominator of the 
running coupling constant (\run) with a universal coefficient. The observation of such a universality demands 
the logarithm to be large enough a large enough such that the first few terms of the expansion represent 
a reasonable approximation. Experimentally, a quasi-two dimensional gas can be implemented with a strongly 
anisotropic trap, which 
can be modeled as a three dimensional gas in a narrow harmonic well in one direction, referred to as the trapped 
direction. An analytical solution to 
the two-body scattering for large extension of the single particle wave function in the trapped direction, $l$, in 
comparison with the range of the inter-particle interaction (which remains three-dimensional) was obtained in 
\ref{\petrov}{D. S. Petrov, M. Holzmann and G. V. Shlyapnikov, Phys. Rev. Lett., {\bf 84}, 2551(2000).}, and an 
effective phase shift of the two dimensional $s$-wave component can be extracted when the de Broglie wavelength 
in the trapped plane becomes much longer than $l$,
$$\delta_{\rm{eff.}}=-{\pi\over 2\ln{1\over ka_{\rm{eff.}}}},\numbereq\name{\eff}$$
where $k$ is the relative 2D momentum and $a_{\rm{eff.}}=\sqrt{\pi}l\exp\Big(-{\sqrt{\pi\over 2}}{l\over a}\Big)$ 
with $a$ the 3D scattering length. The authors of [\petrov] also found numerically that the approximation (\eff) works 
well even when $l\sim a$. The typical inter-particle distance in an atomic trap is about $10^4\AA$ and the typical 3D scattering 
length for alkaline atoms is of the order of $100\AA$. For the thermal wavelength equal to the inter-particle 
distance and the trapped dimension equal to $a$ (which is technically feasible), we have $a_{\rm{eff.}}\simeq
51\AA$ and then $\alpha\simeq0.12$, 
according to (\run) with $a$ there replaced by $a_{\rm{eff.}}$. The universal 2D 
logarithm could be quite significant to observations.
    
\noindent
{\bf Acknowledgment}
I would like to thank Professor N. Khuri for valuable discussions. This work is supported in part by the US 
Department of Energy under grants DE-FG02-91ER40651-TASKB.

\noindent
{\bf Appendix A}

To regularize the infinite volume limit of $V_m(p^\prime, p)$ defined in (\partialw), we follow the 
usual practice by restricting the relative coordinate $\vec r$ within a large circle of radius $R$ and 
imposing the Dirichlet boundary conditions for the wave functions with and without potential $V$, 
i.e. $$u_m(k|R)=0\eqno(A1)$$ and $$u_m^{(0)}(p|R)=0,\eqno(A2)$$
where $u_m(k|r)$ is the solution of the radial Schroedinger equation, regular at $r=0$, 
$$\Big[-{2\over r}{d\over dr}\Big(r{d\over dr}\Big)+{2m^2\over r^2}-V(r)\Big]u_m(k|r)=2k^2u_m(k|r),
\eqno(A3)$$ 
and $u_m^{(0)}(p|r)=J_m(pr)$ satisfies the radial Schroedinger equation of a free particle, i.e.
$$\Big[-{2\over r}{d\over dr}\Big(r{d\over dr}\Big)+{2m^2\over r^2}]u_m^{(0)}(k|r)=2p^2u_m^{(0)}(k|r).
\eqno(A4)$$ 
For $pr>>1$, the asymptotic behavior of $u_m^{(0)}(p|r)$ is 
$$u_m^{(0)}(p|r)\simeq\sqrt{2\over\pi pr}\cos\Big(pr-{m\pi\over 2}-{\pi\over 4}\Big)\eqno(A5)$$ 
and that of $u_m(k|r)$ for $kr>>1$ is given by (\asympt). It follows from (A1), (A2), (A5) and (\asympt)
that
$$p=p_n=\Big(n+{m\over 2}+{1\over 4}\Big){\pi\over R}\eqno(A6)$$
and $$k=k_n=p_n-{\delta_m(p_n)\over R}\eqno(A7)$$ for large $R$.

Multiplying (A3) by $J_m(pr)$ and (A2) by $u_m(k|r)$, subtracting the result and using the boundary 
condition (A1) and (A2), we find
$$\int_0^R drrJ_m(pr)u_m(k|r)={v_m(p,k)\over2(k^2-p^2)},\eqno(A8)$$
where 
$$v_m(p,k)=\int_0^RdrrJ_m(pr)V(r)u_m(k|r).\eqno(A9)$$
and is well behaved in the limit $R\to\infty$. The eq. (\bessel) becomes then
$$V_m(p^\prime,p)=\pi\sum_kN_{km}^2e^{-2\beta k^2}{v_m(p^\prime,k)v_m(p,k)\over k^2-p^2}\eqno(A10)$$
with $N_{km}$ the normalization constant such that
$$N_{km}^2\int_0^Rdrru_m^2(k|r)=1.\eqno(A11)$$
Using the asymptotic behavior (\asympt), we find $N_{km}\simeq\sqrt{{\pi k\over R}}$ for large $R$. 

To take the limit $R\to\infty$ of the sum (A10), we need to isolate out the $k$'s which are sufficiently close to $p$. For this 
purpose,  we introduce a subset of $k$'s, $B$, such that all $k\in B$ satisfy the condition that
$|k-p|<{N\pi\over R}$ for $N>>1$. We further specify the order of the limit such that $R\to\infty$ first and then 
$N\to\infty$. Consequently, the summation in (A10) is divided into two parts,
$$V_m(p^\prime,p)=V_m^<(p^\prime,p)+V_m^>(p^\prime,p).\eqno(A12)$$
with $V_m^<(p^\prime,p)$ including only the $k$'s within $B$ and $V_m^<(p^\prime,p)$ all others.
For $V_m^<(p^\prime,p)$, all the $k$'s except those in the denominator can all be set to $p$, and we obtain
that
$$V_m^<(p^\prime,p)\simeq {\pi^2\over 2}e^{-2\beta p^2}v_m(p^\prime,p)v_m(p,p)\sum_{l=-N}^N{1\over\pi l-\delta_m(p)}
={\pi^2\over 2} e^{-2\beta p^2}v_m(p^\prime,p)v_m(p,p)\cot\delta_m(p).
\eqno(A13)$$
The summation $V_m(p^\prime, p)$, however can be replaced simply by the principal value of an integral, i.e.
$$V_m^>(p^\prime,p)=\pi{\cal P}\int_0^\infty dkke^{-2\beta k^2}{v_m(p^\prime,k)v_m(p,k)\over k^2-p^2}.\eqno(A14)$$

Following the same steps that leads to (A8), we derive that
$$v_m(p,p)=-{4\over\pi}\sin\delta_m(p).\eqno(A15)$$ 

\noindent{\bf Appendix B}

In this appendix, we shall determine the singular behavior of the functions $D(z)$ and $F(z)$ in the Virial
expansion (\eqstate) as $z\to 1^-$.

On writing $D(z)=D_1(z)+D_2(z)$ with
$$D_1(z)\equiv\sum_{r,s=1}^\infty{z^{r+s}\over rs}\ln2rs\eqno(B1)$$
and
$$D_2(z)\equiv-\sum_{r,s=1}^\infty{z^{r+s}\over rs}\ln(r+s),\eqno(B2)$$
we have 
$$D_1(z)=2S_1(z)S_2(z)+S_1^2(z)\ln 2,\eqno(B3)$$ where
$$S_1(z)=\sum_{n=1}^\infty{z^n\over n}=\ln{1\over 1-z}\eqno(B4)$$ and
$$S_2(z)=\sum_{n=1}^\infty{z^n\over n}\ln n\simeq{1\over 2}\ln^2{1\over 1-z},\eqno(B5)$$ 
as $z\to 1^-$. To estimate $D_2(z)$, we rewrite it as 
$$D_2(z)=\sum_{N=1}^\infty C_Nz^N\eqno(B6)$$ where 
$$C_N=\ln N\sum_{m+n=N}{1\over mn}={2\over N}\ln N\sum_{n=1}^{N-1}{1\over n}
\simeq {2\over N}\ln^2N\eqno(B7)$$ for $N>>1$. It follows then that
$$D_2(z)\simeq 2\sum_{N=1}^\infty{z^N\over N}\ln^2N\simeq {2\over 3}\ln^3{1\over 1-z}.\eqno(B8)$$
Combining (B1) and (B2) with the aid of (B3-5) and (B8), we obtain the asymptotic formula
$$D(z)\simeq {1\over 3}\ln^3{1\over 1-z}.\eqno(B9)$$

The asymptotic behavior of $F(z)$ can be determined by its integral representation, read off from 
(\final) and (\res)
$$F(z)=-128\pi^3\beta^2{\cal P}\int{d^2\vec p_1\over (2\pi)^2}\int{d^2\vec p_2\over (2\pi)^2}
\int{d^2\vec p_3\over (2\pi)^2}\int{d^2\vec p_4\over (2\pi)^2}(2\pi)^2\delta^2(\vec p_1+\vec p_2-\vec p_3-\vec p_4)
{n(\vec p_1)n(\vec p_2)n(\vec p_3)\over p_1^2+p_2^2-p_3^2-p_4^2}$$
$$=-32\pi^3\beta^2\int{d^2\vec p_1\over (2\pi)^2}\int{d^2\vec p_2\over (2\pi)^2}
\int{d^2\vec p_3\over (2\pi)^2}\int{d^2\vec p_4\over (2\pi)^2}(2\pi)^2\delta^2(\vec p_1+\vec p_2-\vec p_3-\vec p_4)
$$ $$\times
{n(\vec p_1)n(\vec p_2)[n(\vec p_3)+n(\vec p_4)]-n(\vec p_3)n(\vec p_4)[n(\vec p_1)+n(\vec p_2)]
\over p_1^2+p_2^2-p_3^2-p_4^2}.\eqno(B10)$$
On writing $z=e^{-\epsilon}$, the leading singularity of $F(z)$ in the limit $z\to 1^-$ can be extracted from the 
integration domain where $p_j<{\eta\over\sqrt{\beta}}$ for $j=1,2,3,4$ with $\epsilon<<\eta<<1$. We have then
$$F(z)\simeq 32\pi^3\int_{k_j<\eta}\prod_{j=1}^4{d^2\vec k_j\over (2\pi)^2}{(2\pi)^2\delta^2(\vec k_1+\vec k_2-\vec k_3-
\vec k_4)\over (k_1^2+\epsilon)(k_2^2+\epsilon)(k_3^2+\epsilon)(k_4^2+\epsilon)}$$
$$\simeq{4\over\epsilon}\int_0^\infty dxxK_0^4(x)\eqno(B11)$$
with $K_0(x)$ the modified Bessel function of the second kind.

\noindent
\immediate\closeout1
\bigbreak\bigskip

\line{\twelvebf References. \hfil}
\nobreak\medskip\vskip\parskip

\input refs

\input epsf
\epsfxsize 4truein
\centerline{\epsffile{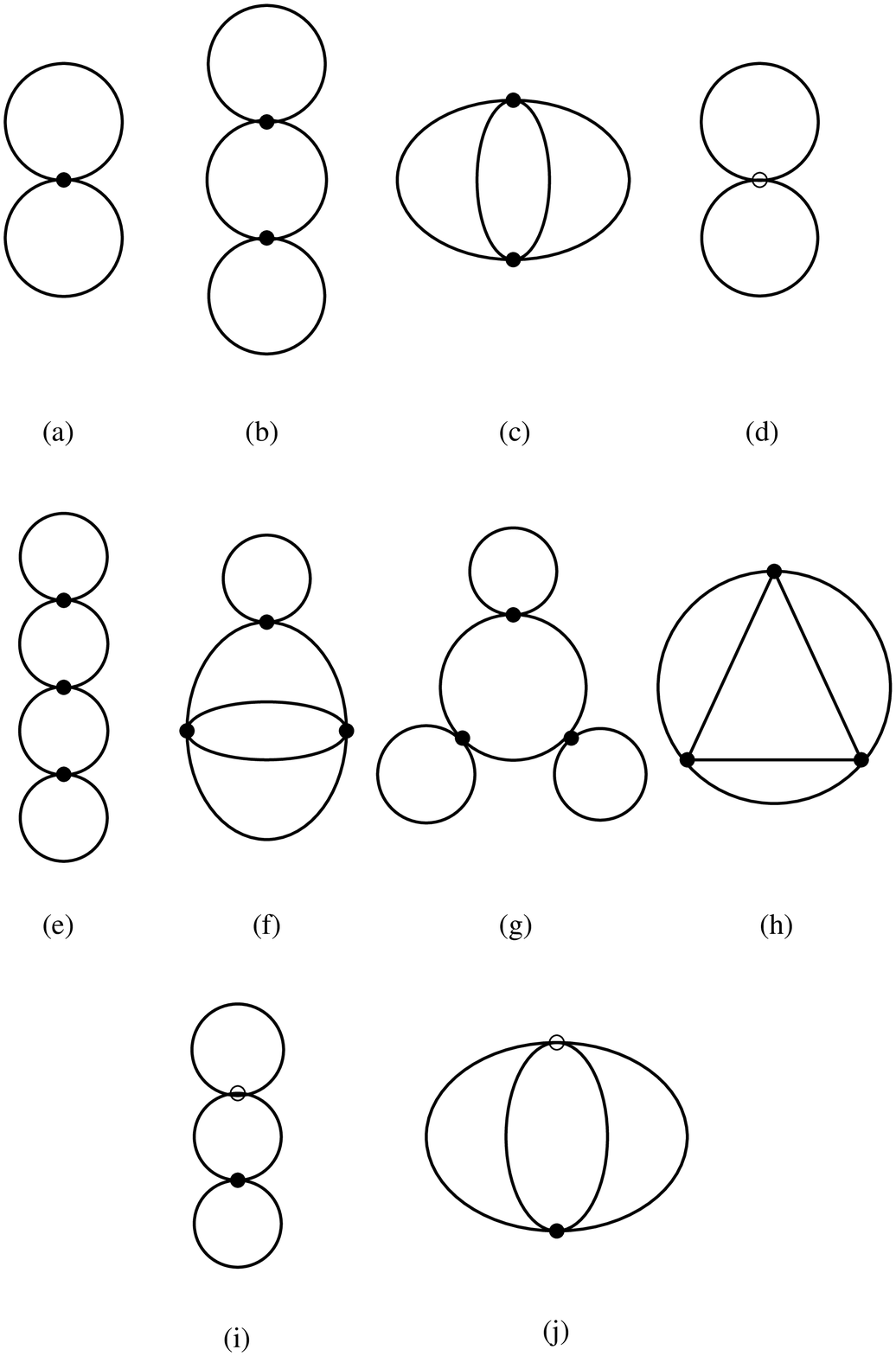}}
\medskip
$$\hbox{Fig. 1 The thermal diagrams for Virial expansion to the order $\alpha^3$}$$

\vfil\end

\bye